\documentclass[zpreprint,zbstpl,british]{zeus_K2.9paper}
\usepackage{zeusdet_units}
\usepackage{lineno}
\usepackage{mwe}
\usepackage{subfig}
\usepackage{pdflscape}
\usepackage{array}
\usepackage{subfig}
\usepackage{epstopdf}
\begin{document}
 \prepnum{DESY--20--048}
 \prepdate{March 2020}

\zeustitle{%
  Study of proton parton distribution functions at high \boldmath$x$ using ZEUS data
}

\zeusauthor{ZEUS Collaboration}
\draftversion{Reading draft}
 \zeusdate{}

\maketitle

%
%
\begin{abstract}\noindent
At large values of $x$ the parton distribution functions (PDFs) of the proton are poorly constrained and there are considerable variations between different global fits. Data at such high $x$ have already been published by the ZEUS Collaboration, but not yet used in PDF extractions.  A technique for comparing predictions based on different PDF sets to the observed number of events in the ZEUS data is presented. It is applied to compare predictions from the most commonly used PDFs to published ZEUS data at high Bjorken $x$. A wide variation is found in the ability of the  PDFs to predict the observed results. A scheme for including the ZEUS high-$x$  data in future PDF extractions is discussed. 
\end{abstract}
\thispagestyle{empty}
\clearpage


%
%
%
%
                                                   %
\begin{center}
{                      \Large  The ZEUS Collaboration              }
\end{center}

{\small\raggedright


I.~Abt$^{19}$, 
L.~Adamczyk$^{7}$, 
R. Aggarwal$^{3, b}$, 
V.~Aushev$^{17}$, 
O.~Behnke$^{9}$, 
U.~Behrens$^{9}$, 
A.~Bertolin$^{21}$, 
I.~Bloch$^{10}$, 
I.~Brock$^{2}$, 
N.H.~Brook$^{28, m}$, 
R.~Brugnera$^{22}$, 
A.~Bruni$^{1}$, 
P.J.~Bussey$^{11}$, 
A.~Caldwell$^{19}$, 
M.~Capua$^{4}$, 
C.D.~Catterall$^{32}$, 
J.~Chwastowski$^{6}$, 
J.~Ciborowski$^{29, o}$, 
R.~Ciesielski$^{9, d}$, 
A.M.~Cooper-Sarkar$^{20}$, 
M.~Corradi$^{1, a}$, 
R.K.~Dementiev$^{18}$, 
S.~Dusini$^{21}$, 
J.~Ferrando$^{9}$, 
B.~Foster$^{20, j}$, 
E.~Gallo$^{13, k}$, 
D.~Gangadharan$^{14}$, 
A.~Garfagnini$^{22}$, 
A.~Geiser$^{9}$, 
L.K.~Gladilin$^{18}$, 
Yu.A.~Golubkov$^{18}$, 
G.~Grzelak$^{29}$, 
C.~Gwenlan$^{20}$, 
D.~Hochman$^{31}$, 
N.Z.~Jomhari$^{9}$, 
I.~Kadenko$^{17}$, 
S.~Kananov$^{23}$, 
U.~Karshon$^{31}$, 
P.~Kaur$^{3, c}$, 
R.~Klanner$^{13}$, 
U.~Klein$^{9, e}$, 
I.A.~Korzhavina$^{18}$, 
N.~Kovalchuk$^{13}$, 
H.~Kowalski$^{9}$, 
O.~Kuprash$^{9, f}$, 
M.~Kuze$^{25}$, 
B.B.~Levchenko$^{18}$, 
A.~Levy$^{23}$, 
B.~L\"ohr$^{9}$, 
A.~Longhin$^{22}$, 
O.Yu.~Lukina$^{18}$, 
I.~Makarenko$^{9}$, 
J.~Malka$^{9, g}$, 
S.~Masciocchi$^{12, i}$, 
K.~Nagano$^{15}$, 
J.D.~Nam$^{24}$, 
J.~Onderwaater$^{14, l}$, 
Yu.~Onishchuk$^{17}$, 
E.~Paul$^{2}$, 
I.~Pidhurskyi$^{17}$, 
A.~Polini$^{1}$, 
M.~Przybycie\'n$^{7}$, 
A.~Quintero$^{24}$, 
M.~Ruspa$^{27}$, 
D.H.~Saxon$^{11}$, 
U.~Schneekloth$^{9}$, 
T.~Sch\"orner-Sadenius$^{9}$, 
I.~Selyuzhenkov$^{12}$, 
M.~Shchedrolosiev$^{17}$, 
L.M.~Shcheglova$^{18}$, 
I.O.~Skillicorn$^{11}$, 
W.~S{\l}omi\'nski$^{8}$, 
A.~Solano$^{26}$, 
L.~Stanco$^{21}$, 
N.~Stefaniuk$^{9}$, 
P.~Stopa$^{6}$, 
B.~Surrow$^{24}$, 
J.~Sztuk-Dambietz$^{13, g}$, 
E.~Tassi$^{4}$, 
K.~Tokushuku$^{15}$, 
M.~Turcato$^{13, g}$, 
O.~Turkot$^{9}$, 
T.~Tymieniecka$^{30}$, 
A.~Verbytskyi$^{19}$, 
W.A.T.~Wan Abdullah$^{5}$, 
K.~Wichmann$^{9}$, 
M.~Wing$^{28, n}$, 
S.~Yamada$^{15}$, 
Y.~Yamazaki$^{16}$, 
A.F.~\.Zarnecki$^{29}$, 
L.~Zawiejski$^{6}$, 
O.~Zenaiev$^{9, h}$ 
\newpage


{\setlength{\parskip}{0.4em}
\makebox[3ex]{$^{1}$}
\begin{minipage}[t]{14cm}
{\it INFN Bologna, Bologna, Italy}~$^{A}$

\end{minipage}

\makebox[3ex]{$^{2}$}
\begin{minipage}[t]{14cm}
{\it Physikalisches Institut der Universit\"at Bonn,
Bonn, Germany}~$^{B}$

\end{minipage}

\makebox[3ex]{$^{3}$}
\begin{minipage}[t]{14cm}
{\it Panjab University, Department of Physics, Chandigarh, India}

\end{minipage}

\makebox[3ex]{$^{4}$}
\begin{minipage}[t]{14cm}
{\it Calabria University,
Physics Department and INFN, Cosenza, Italy}~$^{A}$

\end{minipage}

\makebox[3ex]{$^{5}$}
\begin{minipage}[t]{14cm}
{\it National Centre for Particle Physics, Universiti Malaya, 50603 Kuala Lumpur, Malaysia}~$^{C}$

\end{minipage}

\makebox[3ex]{$^{6}$}
\begin{minipage}[t]{14cm}
{\it The Henryk Niewodniczanski Institute of Nuclear Physics, Polish Academy of \\
Sciences, Krakow, Poland}

\end{minipage}

\makebox[3ex]{$^{7}$}
\begin{minipage}[t]{14cm}
{\it AGH University of Science and Technology, Faculty of Physics and Applied Computer
Science, Krakow, Poland}

\end{minipage}

\makebox[3ex]{$^{8}$}
\begin{minipage}[t]{14cm}
{\it Department of Physics, Jagellonian University, Krakow, Poland}~$^{D}$

\end{minipage}

\makebox[3ex]{$^{9}$}
\begin{minipage}[t]{14cm}
{\it Deutsches Elektronen-Synchrotron DESY, Hamburg, Germany}

\end{minipage}

\makebox[3ex]{$^{10}$}
\begin{minipage}[t]{14cm}
{\it Deutsches Elektronen-Synchrotron DESY, Zeuthen, Germany}

\end{minipage}

\makebox[3ex]{$^{11}$}
\begin{minipage}[t]{14cm}
{\it School of Physics and Astronomy, University of Glasgow,
Glasgow, United Kingdom}~$^{E}$

\end{minipage}

\makebox[3ex]{$^{12}$}
\begin{minipage}[t]{14cm}
{\it GSI Helmholtzzentrum f\"{u}r Schwerionenforschung GmbH, Darmstadt, Germany}

\end{minipage}

\makebox[3ex]{$^{13}$}
\begin{minipage}[t]{14cm}
{\it Hamburg University, Institute of Experimental Physics, Hamburg,
Germany}~$^{F}$

\end{minipage}

\makebox[3ex]{$^{14}$}
\begin{minipage}[t]{14cm}
{\it Physikalisches Institut of the University of Heidelberg, Heidelberg, Germany}

\end{minipage}

\makebox[3ex]{$^{15}$}
\begin{minipage}[t]{14cm}
{\it Institute of Particle and Nuclear Studies, KEK,
Tsukuba, Japan}~$^{G}$

\end{minipage}

\makebox[3ex]{$^{16}$}
\begin{minipage}[t]{14cm}
{\it Department of Physics, Kobe University, Kobe, Japan}~$^{G}$

\end{minipage}

\makebox[3ex]{$^{17}$}
\begin{minipage}[t]{14cm}
{\it Department of Nuclear Physics, National Taras Shevchenko University of Kyiv, Kyiv, Ukraine}

\end{minipage}

\makebox[3ex]{$^{18}$}
\begin{minipage}[t]{14cm}
{\it Lomonosov Moscow State University, Skobeltsyn Institute of Nuclear Physics,
Moscow, Russia}

\end{minipage}

\makebox[3ex]{$^{19}$}
\begin{minipage}[t]{14cm}
{\it Max-Planck-Institut f\"ur Physik, M\"unchen, Germany}

\end{minipage}

\makebox[3ex]{$^{20}$}
\begin{minipage}[t]{14cm}
{\it Department of Physics, University of Oxford,
Oxford, United Kingdom}~$^{E}$

\end{minipage}

\makebox[3ex]{$^{21}$}
\begin{minipage}[t]{14cm}
{\it INFN Padova, Padova, Italy}~$^{A}$

\end{minipage}

\makebox[3ex]{$^{22}$}
\begin{minipage}[t]{14cm}
{\it Dipartimento di Fisica e Astronomia dell' Universit\`a and INFN,
Padova, Italy}~$^{A}$

\end{minipage}

\makebox[3ex]{$^{23}$}
\begin{minipage}[t]{14cm}
{\it Raymond and Beverly Sackler Faculty of Exact Sciences, School of Physics, \\
Tel Aviv University, Tel Aviv, Israel}~$^{H}$

\end{minipage}

\makebox[3ex]{$^{24}$}
\begin{minipage}[t]{14cm}
{\it Department of Physics, Temple University, Philadelphia, PA 19122, USA}~$^{I}$

\end{minipage}

\makebox[3ex]{$^{25}$}
\begin{minipage}[t]{14cm}
{\it Department of Physics, Tokyo Institute of Technology,
Tokyo, Japan}~$^{G}$

\end{minipage}

\makebox[3ex]{$^{26}$}
\begin{minipage}[t]{14cm}
{\it Universit\`a di Torino and INFN, Torino, Italy}~$^{A}$

\end{minipage}

\makebox[3ex]{$^{27}$}
\begin{minipage}[t]{14cm}
{\it Universit\`a del Piemonte Orientale, Novara, and INFN, Torino,
Italy}~$^{A}$

\end{minipage}

\makebox[3ex]{$^{28}$}
\begin{minipage}[t]{14cm}
{\it Physics and Astronomy Department, University College London,
London, United Kingdom}~$^{E}$

\end{minipage}

\makebox[3ex]{$^{29}$}
\begin{minipage}[t]{14cm}
{\it Faculty of Physics, University of Warsaw, Warsaw, Poland}

\end{minipage}

\makebox[3ex]{$^{30}$}
\begin{minipage}[t]{14cm}
{\it National Centre for Nuclear Research, Warsaw, Poland}

\end{minipage}

\makebox[3ex]{$^{31}$}
\begin{minipage}[t]{14cm}
{\it Department of Particle Physics and Astrophysics, Weizmann
Institute, Rehovot, Israel}

\end{minipage}

\makebox[3ex]{$^{32}$}
\begin{minipage}[t]{14cm}
{\it Department of Physics, York University, Ontario, Canada M3J 1P3}~$^{J}$

\end{minipage}

}

\vspace{3em}


{\setlength{\parskip}{0.4em}\raggedright
\makebox[3ex]{$^{ A}$}
\begin{minipage}[t]{14cm}
 supported by the Italian National Institute for Nuclear Physics (INFN) \
\end{minipage}

\makebox[3ex]{$^{ B}$}
\begin{minipage}[t]{14cm}
 supported by the German Federal Ministry for Education and Research (BMBF), under
 contract No.\ 05 H09PDF\
\end{minipage}

\makebox[3ex]{$^{ C}$}
\begin{minipage}[t]{14cm}
 supported by HIR grant UM.C/625/1/HIR/149 and UMRG grants RU006-2013, RP012A-13AFR and RP012B-13AFR from
 Universiti Malaya, and ERGS grant ER004-2012A from the Ministry of Education, Malaysia\
\end{minipage}

\makebox[3ex]{$^{ D}$}
\begin{minipage}[t]{14cm}
supported by the Polish National Science Centre (NCN) grant no.\ DEC-2014/13/B/ST2/02486
\end{minipage}

\makebox[3ex]{$^{ E}$}
\begin{minipage}[t]{14cm}
 supported by the Science and Technology Facilities Council, UK\
\end{minipage}

\makebox[3ex]{$^{ F}$}
\begin{minipage}[t]{14cm}
 supported by the German Federal Ministry for Education and Research (BMBF), under
 contract No.\ 05h09GUF, and the SFB 676 of the Deutsche Forschungsgemeinschaft (DFG) \
\end{minipage}

\makebox[3ex]{$^{ G}$}
\begin{minipage}[t]{14cm}
 supported by the Japanese Ministry of Education, Culture, Sports, Science and Technology
 (MEXT) and its grants for Scientific Research\
\end{minipage}

\makebox[3ex]{$^{ H}$}
\begin{minipage}[t]{14cm}
 supported by the Israel Science Foundation\
\end{minipage}

\makebox[3ex]{$^{ I}$}
\begin{minipage}[t]{14cm}
 supported in part by the Office of Nuclear Physics within the U.S.\ DOE Office of Science
\end{minipage}

\makebox[3ex]{$^{ J}$}
\begin{minipage}[t]{14cm}
 supported by the Natural Sciences and Engineering Research Council of Canada (NSERC) \
\end{minipage}

}

\pagebreak[4]
{\setlength{\parskip}{0.4em}


\makebox[3ex]{$^{ a}$}
\begin{minipage}[t]{14cm}
now at INFN Roma, Italy\
\end{minipage}

\makebox[3ex]{$^{ b}$}
\begin{minipage}[t]{14cm}
now at DST-Inspire Faculty, Department of Technology, SPPU, India\
\end{minipage}

\makebox[3ex]{$^{ c}$}
\begin{minipage}[t]{14cm}
now at Sant Longowal Institute of Engineering and Technology, Longowal, Punjab, India\
\end{minipage}

\makebox[3ex]{$^{ d}$}
\begin{minipage}[t]{14cm}
now at Rockefeller University, New York, NY 10065, USA\
\end{minipage}

\makebox[3ex]{$^{ e}$}
\begin{minipage}[t]{14cm}
now at University of Liverpool, United Kingdom\
\end{minipage}

\makebox[3ex]{$^{ f}$}
\begin{minipage}[t]{14cm}
now at University of Freiburg, Freiburg, Germany\
\end{minipage}

\makebox[3ex]{$^{ g}$}
\begin{minipage}[t]{14cm}
now at European X-ray Free-Electron Laser facility GmbH, Hamburg, Germany\
\end{minipage}

\makebox[3ex]{$^{ h}$}
\begin{minipage}[t]{14cm}
now at Hamburg University, Hamburg, Germany\
\end{minipage}

\makebox[3ex]{$^{ i}$}
\begin{minipage}[t]{14cm}
also at Physikalisches Institut of the University of Heidelberg, Heidelberg,  Germany\
\end{minipage}

\makebox[3ex]{$^{ j}$}
\begin{minipage}[t]{14cm}
also at DESY and University of Hamburg\
\end{minipage}

\makebox[3ex]{$^{ k}$}
\begin{minipage}[t]{14cm}
also at DESY\
\end{minipage}

\makebox[3ex]{$^{ l}$}
\begin{minipage}[t]{14cm}
also at GSI Helmholtzzentrum f\"{u}r Schwerionenforschung GmbH, Darmstadt, Germany\
\end{minipage}

\makebox[3ex]{$^{ m}$}
\begin{minipage}[t]{14cm}
now at University of Bath, United Kingdom\
\end{minipage}

\makebox[3ex]{$^{ n}$}
\begin{minipage}[t]{14cm}
also supported by DESY\
\end{minipage}

\makebox[3ex]{$^{ o}$}
\begin{minipage}[t]{14cm}
also at Lodz University, Poland\
\end{minipage}

}

}

\clearpage
\pagenumbering{arabic}
%
%

\section{Introduction}
\label{sec-int}

Important questions related to the nature of the strong interaction can be addressed by studying the structure of particles composed of quarks and gluons.  The spatial and momentum distributions of these constituents in a hadron are not understood from first principles owing to theoretical and calculational limitations.  Thus, our knowledge stems principally from measurements.  

Knowledge of the proton structure is also of great importance, as protons are used to reach the high-energy frontier in particle colliders such as the Large Hadron Collider (LHC).  Calculated cross sections for processes involving protons are based on sets of parton distribution functions (PDFs), i.e.,~quark and gluon distribution functions.  These PDFs have been extracted by a number of collaborations by selecting data sets from various experiments. The PDF sets used for making predictions for LHC processes, and a prescription for how uncertainties related to the PDFs are to be evaluated, have been published elsewhere~\cite{PDFsummary:1,PDFsummary:2}.  
Identifying effects due to new physics beyond the Standard Model at high center-of-mass energies requires precise knowledge of the PDFs at high $x$, where $x$ is the fractional longitudinal momentum of the struck parton inside the proton. This knowledge is essential for the isolation of some types of new physics. 


Most of the data available for $ x \geq 0.6$ were obtained in fixed-target experiments
in a range of $Q^2$, the negative 4-momentum-transfer squared, where perturbative quantum chromodynamics (pQCD) may not be fully applicable.  The HERA (Hadron$-$Elektron Ring Anlage) storage ring, where $920$~GeV protons collided with $27.5$~GeV electrons or positrons, offered an opportunity to probe the region of high Bjorken\footnote{From here on, no distinction is made between $x$ and Bjorken $x$.} $x$ at high $Q^2$ where pQCD and the Dokshitzer--Gribov--Lipatov--Altarelli--Parisi (DGLAP) evolution dynamics~\cite{Gribov:1972ri,Gribov:1972rt,Lipatov:1974qm,Dokshitzer:1977sg,Altarelli:1977zs} are believed to be reliable. The ZEUS collaboration published high-$x$  data collected in the period $2004- 2007$, with integrated luminosities of $187$~pb$^{-1}$ for $e^-p$ and $142$~pb$^{-1}$ for $e^+p$ scattering~\cite{highxpaper:1}. The data cover the $Q^2$ range of $650-20000$~GeV$^2$ and the $x$ range of $0.03-1.0$.  

In the following, predictions based on the principal PDF sets used at the LHC are compared to these ZEUS high-$x$ data. While there is some overlap with data used in other ZEUS publications, a substantial fraction of the high-$x$ data from ZEUS has not been previously used in the extraction of PDFs. Most PDF extractions have used the combined HERA data~\cite{herapdf2.0} for their extraction procedures.  The combined HERA data is reported as cross sections which are extracted from a combination of ZEUS and H1 data.  The combination was based on an agreed binning of the kinematically allowed space.  The data discussed in this paper uses much finer binning than used in the data combination and extends the reported data to $x=1$.  Both of these difference lead to bins with small event counts where standard chi-squared fitting is not appropriate and the use of the Poisson distribution to properly account for statistical fluctuations is required. In this paper, the predictions for the expected numbers of events from the different PDF sets are compared to the numbers observed in the ZEUS data and the results are discussed.  Section~6 of this paper discusses the high-$x$ points published previously~\cite{highxpaper:1} that have overlap with HERA-ZEUS legacy data and procedures for taking the data overlap into account in future PDF extractions.

\section{Method to test high-\boldmath$x$ predictions}
\label{sec-exp}

In addition to reduced cross sections, the previous ZEUS publication~\cite{highxpaper:1} provides the observed numbers of events in  $(\Delta x, \Delta Q^2)$ intervals (`bins').   The evaluation of the probability to have observed these numbers of events according to different PDF sets is discussed in the present paper.  The probability, $P(D|\mathrm{PDF}_k)$, for a prediction based on a given PDF set $k$ to predict the data set $D$ is

\begin{equation}
\label{eq:probability}
P(D|\mathrm{PDF_k}) = \prod_j \frac{e^{-\nu_{j,k}} \nu_{j,k}^{n_j}}{n_j !},
\end{equation}
where the index $j$ labels the bins in $(\Delta x, \Delta Q^2)$, $\nu_{j,k}$ is the expected number of events in bin $j$ as predicted from PDF set $k$  and  $n_j$ is the observed number of events. The effect of systematic uncertainties is evaluated by varying the predictions $\nu_{j,k}$ according to the sources of systematic uncertainty as described in Section~\ref{sec:systematics} below. 

 The  probability values in Eq.~(\ref{eq:probability}) are very small absolute quantities and only the ratio of probabilities for the different PDF sets are of interest; i.e., the Bayes Factors.  Using these, an effective $\Delta \chi^2$ between two different PDF sets, here labeled $k,l$,  is defined via
\begin{equation}
\Delta \chi^2_{k,l} = -2 \ln \frac{P(D|\mathrm{PDF}_k)}{P(D|\mathrm{PDF}_l)} \; .
\label{Eq:chsq}
\end{equation}
These quantities are used to evaluate the relative goodness-of-fit of the different PDF sets.  The tail-area probabilities ($p$-values~\cite{Pvalues:1}) for the different PDF sets are also provided, based on the expected probability distribution of the quantities $P(D|\mathrm{PDF_{\it{k}}})$.   These are used to evaluate the overall goodness-of-fit for the prediction based on the PDF set.


To calculate the probabilities as outlined above, the predictions from the PDF sets must be
evaluated.  The PDFs are used as input to calculate cross sections for the $e^{\pm}p \rightarrow e^{\pm}X$ neutral current cross sections at the  Born level with fixed fine structure constant, $\alpha(Q^2=0)$.  The predictions for the observed number of events in measured kinematic variables, $(x_{\rm rec}, Q^{2}_{\rm rec} )$, are given by integrating over the full kinematic phase space:
\begin{equation}
 \nu_{j,k} =\mathcal{L} \int_{(\Delta x,\Delta Q^{2})_j} \left[\int A(x_{\rm rec},Q^{2}_{\rm rec}|x,Q^{2}) \frac{d^2 \sigma(x,Q^{2}|\mathrm{PDF}_k)}{dx dQ^2}  dxdQ^{2} \right]   dx_{\rm rec}dQ^{2}_{\rm rec} .
\label{eqn:nu_integ}		
\end{equation}
Here, $\mathcal{L}$ is the luminosity, $\frac{d^2 \sigma(x,Q^{2}|\mathrm{PDF}_k)}{dx dQ^2}$ is the differential cross section at $(x,Q^{2})$ for PDF set $k$ using kinematic quantities defined at the Born level and $A(x_{\rm rec},Q^{2}_{\rm rec}|x,Q^{2}) $ transforms the Born level cross sections to observed cross sections  including all relevant effects (radiative corrections, detector resolution and acceptance, selection criteria, etc.).  This integral is approximated as 
\begin{equation}
\nu_{j,k} \approx \sum_{i} A_{ji}\lambda_{i,k}  \; ,
\label{eqn:nu_j}		
\end{equation}
where $\lambda_{i,k}$ is the expected number of events for the $i^{\rm th}$  $(\Delta x,\Delta Q^{2})$ bin at the Born level for PDF set $k$, and $A_{ji}$ gives the transformation to the expectation in the measured quantities in bin $j$. 
In the following,  $\boldsymbol \lambda_k$ represent the vectors $\{\lambda_{i,k}\}$ and  $\boldsymbol \nu_k$ the $\{\nu_{j,k}\}$.
   For the prediction in Eq.~(\ref{eqn:nu_j}) to be accurate and sufficiently precise, 
 it is necessary to have sufficiently fine binning, and to account for all effects through  reliable simulation packages. Thus, the bins defined at the Born level are not identical to those used in the measurements; the index $i$ spans a larger range than the index $j$.
 
 The matrix $\boldsymbol A$ (with components $A_{ji}$)  was decomposed into a matrix representing the QED and QCD radiative effects, $\boldsymbol R$, and a matrix, $\boldsymbol T$, representing the ZEUS specific detector effects:
\begin{equation}
\boldsymbol A = \boldsymbol T\boldsymbol R \; .		
\end{equation}

 The effects of QED and QCD radiative corrections, including the running of the coupling constants, were determined by running  dedicated simulation programs which provide predictions at the `generator level'.  The matrix $\boldsymbol R$ transforms the Born-level quantities, $\lambda_{i,k} $, to the generator-level quantities, $ \mu_{l,k} $:
\begin{equation}
  \mu_{l,k} = \sum_i R_{li}\lambda_{i,k} \; ,
\label{eqn:a_ij}		
\end{equation}
 \noindent where the same binning was used for $ \mu_{l,k}$ and $\lambda_{i,k}$.
 
The matrix $\boldsymbol T$ provides the transformation of the generator-level quantities to the observed quantities.  It accounts for all experimental and analysis-related effects, and is, in principle, independent of the PDF set used in the generation of the Monte Carlo simulated events.

Given the notation above, for a given PDF set $k$, the vector ${\boldsymbol \nu_k}$ is given as
\begin{equation}
{\boldsymbol \nu}_k ={\boldsymbol T} {\boldsymbol \mu}_k \; ,	
\end{equation}
where 
\begin{equation}
{\boldsymbol \mu}_k = {\boldsymbol R \boldsymbol \lambda}_k \; .		
\end{equation}

\section{Evaluation of $\boldsymbol R$ and $\boldsymbol T$}
\subsection{Monte Carlo samples}
The event simulation was performed using the HERACLES~\cite{cpc:69:155} event generator together with the NLO CTEQ5D~\cite{epj:c12:375} PDF set, which is extracted using the zero-mass convention for heavy flavors. The HERACLES program  applies $O(\alpha)$ QED and $O(\alpha_s)$ QCD corrections, where $\alpha_s$ is strong coupling constant.    The QCD simulation is based on a combination of ARIADNE4.12~\cite{cpc:71:15} and MEPS6.5.1~\cite{cpc:101:108} parton-shower simulations as used in the ZEUS high-$x$ analysis~\cite{highxpaper:1}.  The detector simulation was carried out using GEANT3.21~\cite{tech:cern-dd-ee-84-1} tuned to match the ZEUS detector performance.  Exactly the same analysis steps that were carried out in the extraction of the observed event numbers~\cite{highxpaper:1} were applied in the extraction of the Monte Carlo (MC) event numbers. 

The MC events were produced in sets with different minimum $Q^2$ values in order to have sufficient statistical precision at the highest $x$ and $Q^2$ values.  The  MC event sets were then combined to produce the final sample.  This necessitated the introduction of MC event weights, $\omega^{\rm MC}$, to ensure the correct overall cross section, $ \frac{d^2 \sigma(x,Q^{2})}{dx dQ^2}$, as a function of the kinematic variables.  Event weights were also introduced in order to improve the MC description of various experimental quantities such as the event-vertex distribution and selection-cut efficiencies~\cite{highxpaper:1}.  The product of these weights is denoted as $\omega^{\rm sim}$.  The total weight for an event $m$ is then $\omega_m = \omega^{\rm MC}_{m} \omega^{\rm sim}_{m}$.


\subsection{Determination of matrix elements}

The ZEUS analysis~\cite{highxpaper:1} used 153 $(\Delta x, \Delta Q^2)$ bins, fixing the size of the vector $\boldsymbol \nu_k$.  The bins for the  event-generator quantities were chosen on a finer scale, and also extend beyond the kinematic region covered in the measured variables in order to capture all `migrations' from the generated to the measured quantities.  Stable results for the predictions were obtained by choosing 429 bins for the $\boldsymbol \lambda_k$ and $\boldsymbol \mu_k$ vectors, so that the matrix $\boldsymbol T$  has  $153 \times 429$ entries. In particular, it was verified that the predictions for the expected event numbers $\boldsymbol \nu_{k=\rm CTEQ5D}$ as found in the original analysis~\cite{highxpaper:1} were reproduced.

The matrix $\boldsymbol R$ is a diagonal $429 \times 429$ matrix whose elements were evaluated as
\begin{equation}
 R_{ii} = \frac{\sum^{M_{i}}_{m=1}\omega_{m}^{\rm{MC}} }{\mathcal {L^{\rm MC}} \sigma_{i,{\rm CTEQ5D}}}  
 \end{equation}
\noindent where $M_{i}$ is the number of  generated events in the $i^{\rm th}$  $(\Delta x,\Delta Q^{2})$  bin, with the kinematic quantities calculated using the four-vectors of the exchanged boson and incoming proton, and $\mathcal{L}^{\rm MC}$ and $\sigma_{i,{\rm CTEQ5D}}$ are the MC luminosity and Born-level cross section for bin $i$. Non-diagonal elements of ${\boldsymbol R}$ were set to zero.  It was verified that this simplified treatment of the radiative corrections is adequate for the purposes of this analysis and that $\boldsymbol R$ does not depend on the PDF set.

\begin{figure}
\centering
\subfloat[]{\includegraphics[scale=.65]{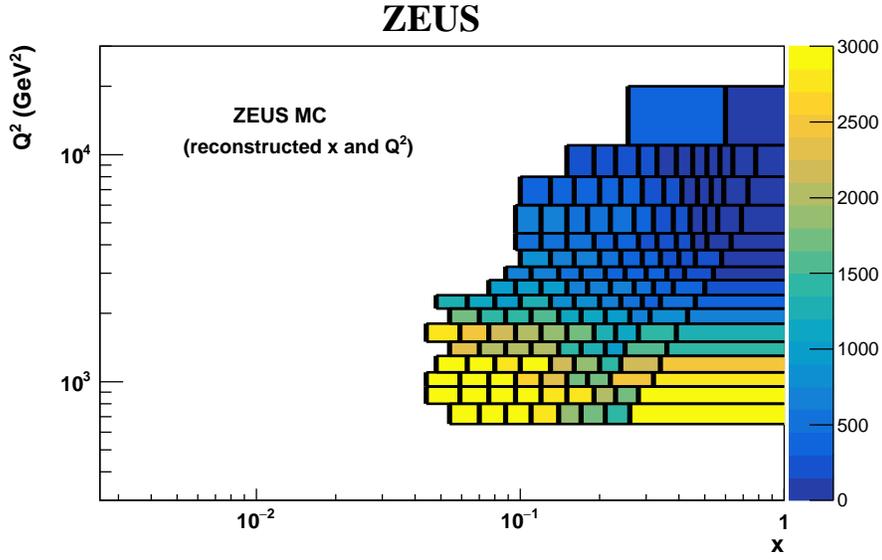}}\\
\subfloat[]{\includegraphics[scale=.65]{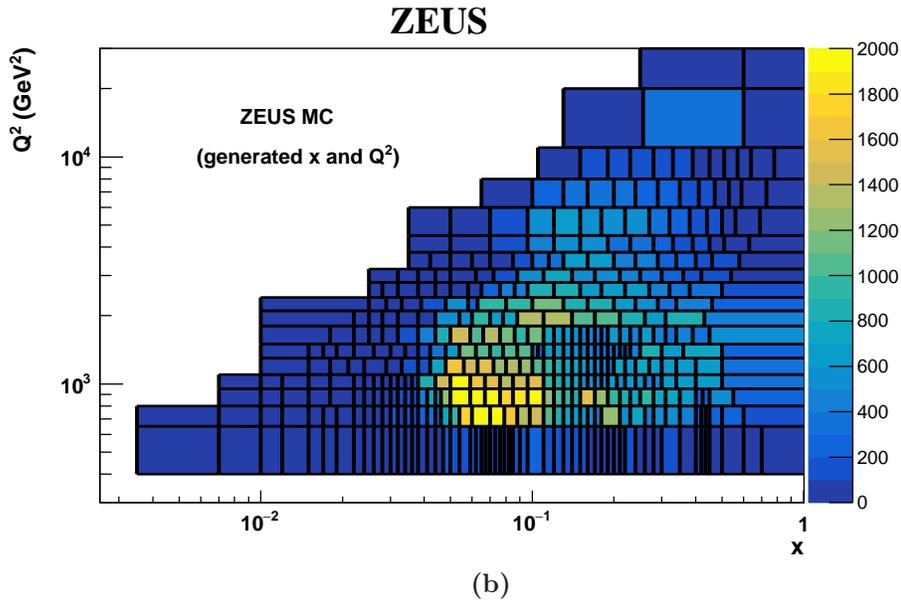}}
\caption{ (a) Distribution of the expected event numbers $\boldsymbol\nu_k$ in the Monte Carlo simulation after all analysis selections are applied, shown in bins of the measured quantities. (b) Distribution of the same events, shown in bins of the generated kinematic quantities $\boldsymbol\mu_k$.  }
\label{fig:Rec_events_bins}
\end{figure}

The elements of $\boldsymbol T$  were calculated as
\begin{equation}
T_{ji} = \frac{\sum^{M_{i}}_{m=1} \omega_{m} I (m \in j)}{\sum^{M_{i}}_{m=1} \omega_{m}^{\rm {MC}}}  \; .
\label{eqn:a_ij}		
\end{equation}
The indicator function $I(m \in j) = 1$ if event $m$ is reconstructed in bin $j$, else $I(m \in j) = 0$.
Figure~\ref{fig:Rec_events_bins} (a) shows the distribution of $\boldsymbol {\nu}_{k=\rm {CTEQ5D}}$ in bins of the measured kinematic quantities.  Figure~\ref{fig:Rec_events_bins} (b) shows the distribution of the same events in the $\boldsymbol \mu_k$ binning.  
\section{Results}
\subsection{Comparison of predictions based on different PDF sets}

\begin{landscape}
\begin{figure}
\centering
\includegraphics[scale=.98]{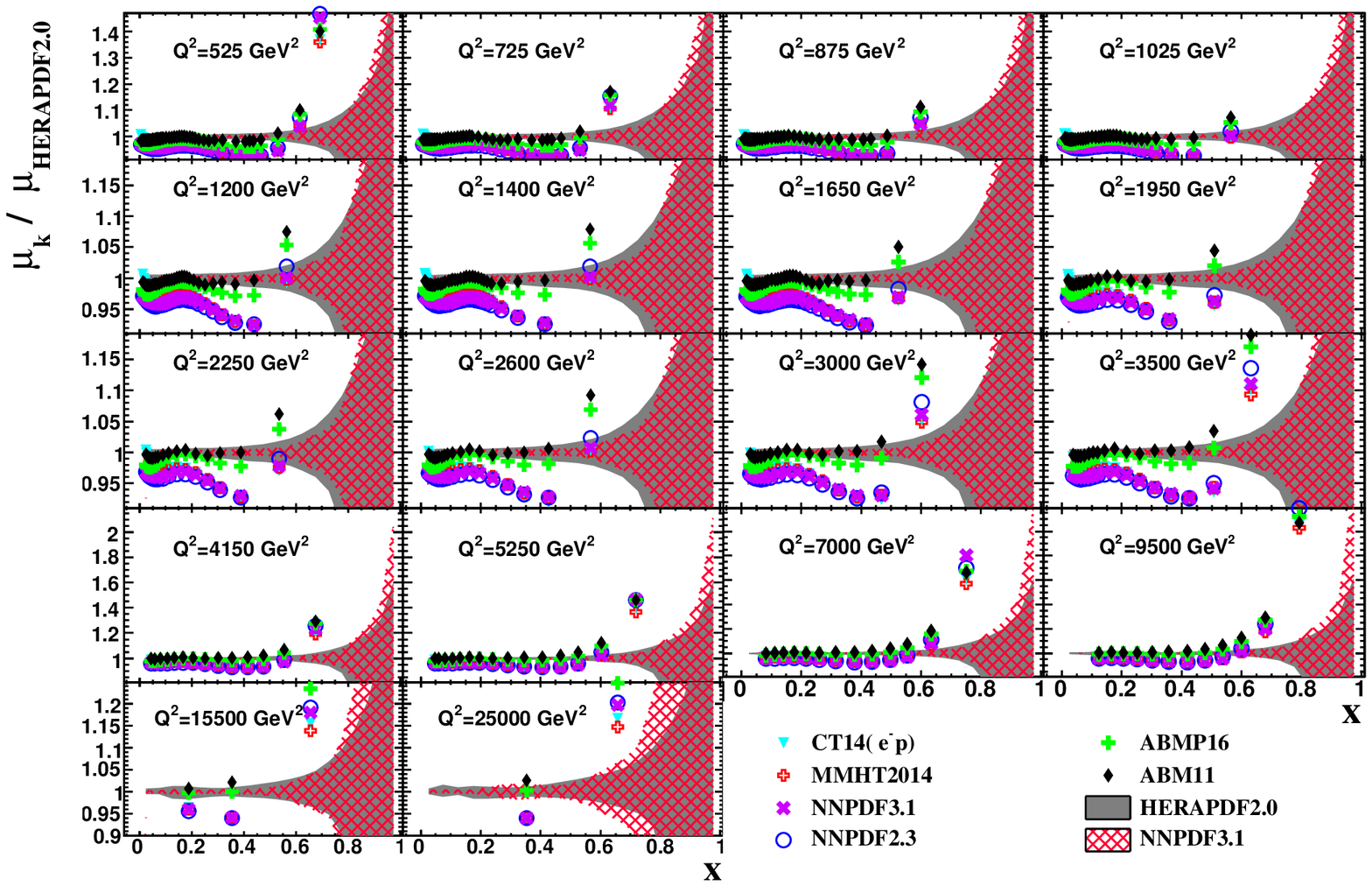}
\caption{Ratios of the  elements of $\boldsymbol \mu_k$ for the PDF sets listed to those calculated using HERAPDF2.0 as functions of $x$ in different $Q^2$ intervals for $e^-p$ data.  The  $Q^2$  value is given at the center of the bin, while the $x$ value is at the mean $x$ value of the events predicted in the bin for HERAPDF2.0.  The shaded band represents the uncertainty quoted by HERAPDF2.0.  The cross-hatched band represents the uncertainty quoted by NNPDF3.1.}
\label{fig:eMGen_events_bins}
\end{figure}
\end{landscape}
\begin{landscape}
\begin{figure}
\centering
\includegraphics[scale=.98]{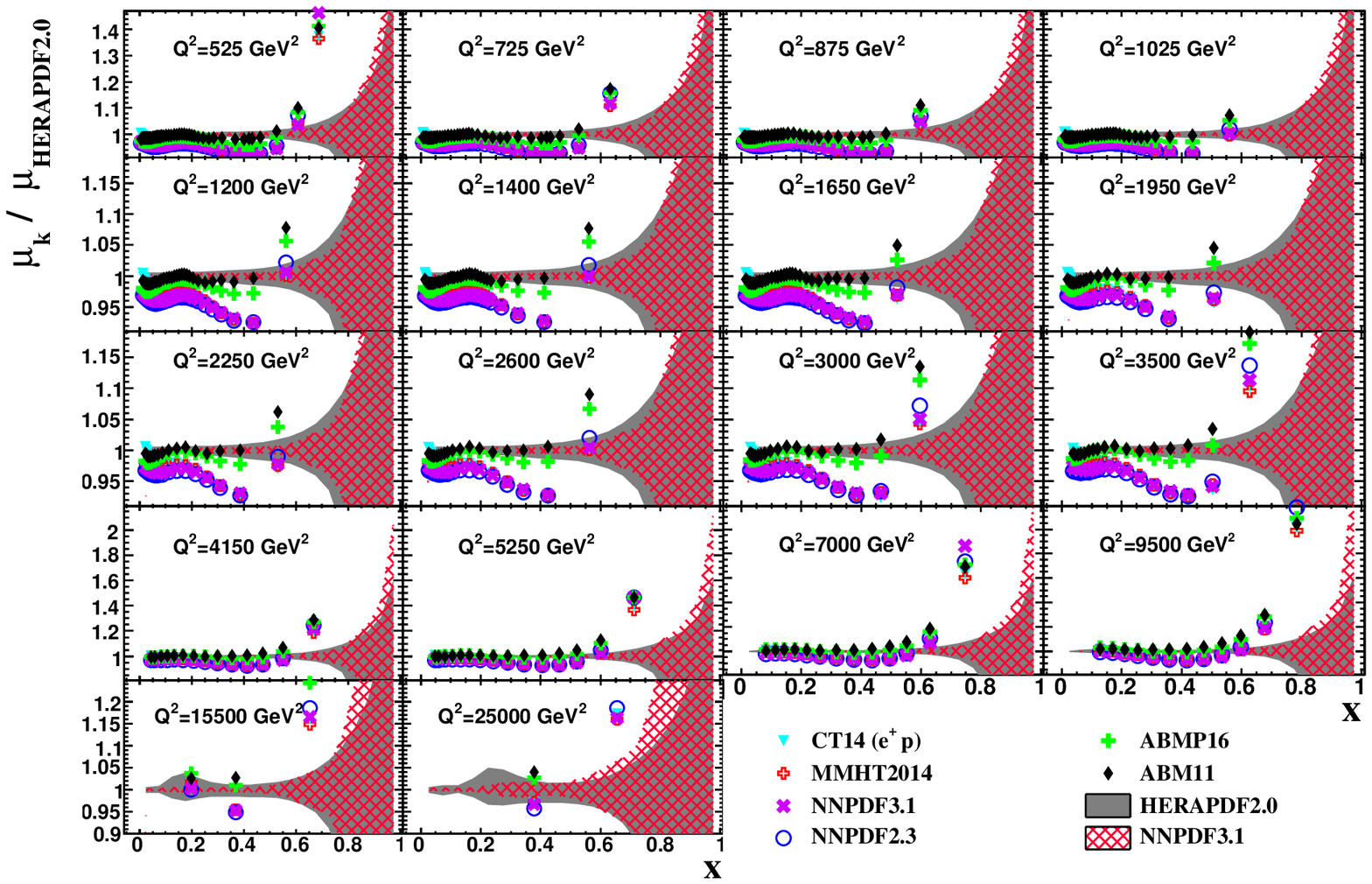}
\caption{Ratios of the  elements of $\boldsymbol \mu_k$ for the PDF sets listed to those calculated using HERAPDF2.0 as functions of $x$ in different $Q^2$ intervals for $e^+p$ data.  The  $Q^2$  value is given at the center of the bin, while the $x$ value is at the mean $x$ value of the events predicted in the bin for HERAPDF2.0.  The shaded band represents the uncertainty quoted by HERAPDF2.0.  The cross-hatched band represents the uncertainty quoted by NNPDF3.1.}\label{fig:ePGen_events_bins}
\end{figure}
\end{landscape}

Figures~\ref{fig:eMGen_events_bins} and~\ref{fig:ePGen_events_bins}  show the ratios of the  elements of ${\boldsymbol \mu}_k$ to the elements of $\boldsymbol {\mu}_{k=\rm {HERAPDF2.0}}$ as functions of $x$ in different $Q^2$ intervals for selected  PDF sets (CT14~\cite{CT:1}, MMHT2014~\cite{MMHT:2}, NNPDF3.1~\cite{Ball:2017}, NNPDF2.3~\cite{Ball:2013}, ABMP16~\cite{abmp16} and ABM11~\cite{ABM:3}). 
The ratio
is shown at the average $x$ value from HERAPDF2.0~\cite{herapdf2.0}  in the bins\footnote{The values of $x$ were calculated for the MC sample from the four-vector of the exchanged boson and the proton.}.  The value of $Q^2$ is for the center of the respective bin.  The expected number of events from the different PDF sets differ by more than a factor of $2$ at large $x$ values.  There is also a systematic difference of several percent in the expectations at the smaller values of $x$ amongst the different sets, which also predict different shapes as a function of $x$. The differences between the PDF predictions significantly exceed the one-standard-deviation uncertainties given by both HERAPDF2.0 and NNPDF3.1.  



\subsection{Comparison of predictions to data}

\begin{figure}
\centering
\includegraphics[scale=.8]{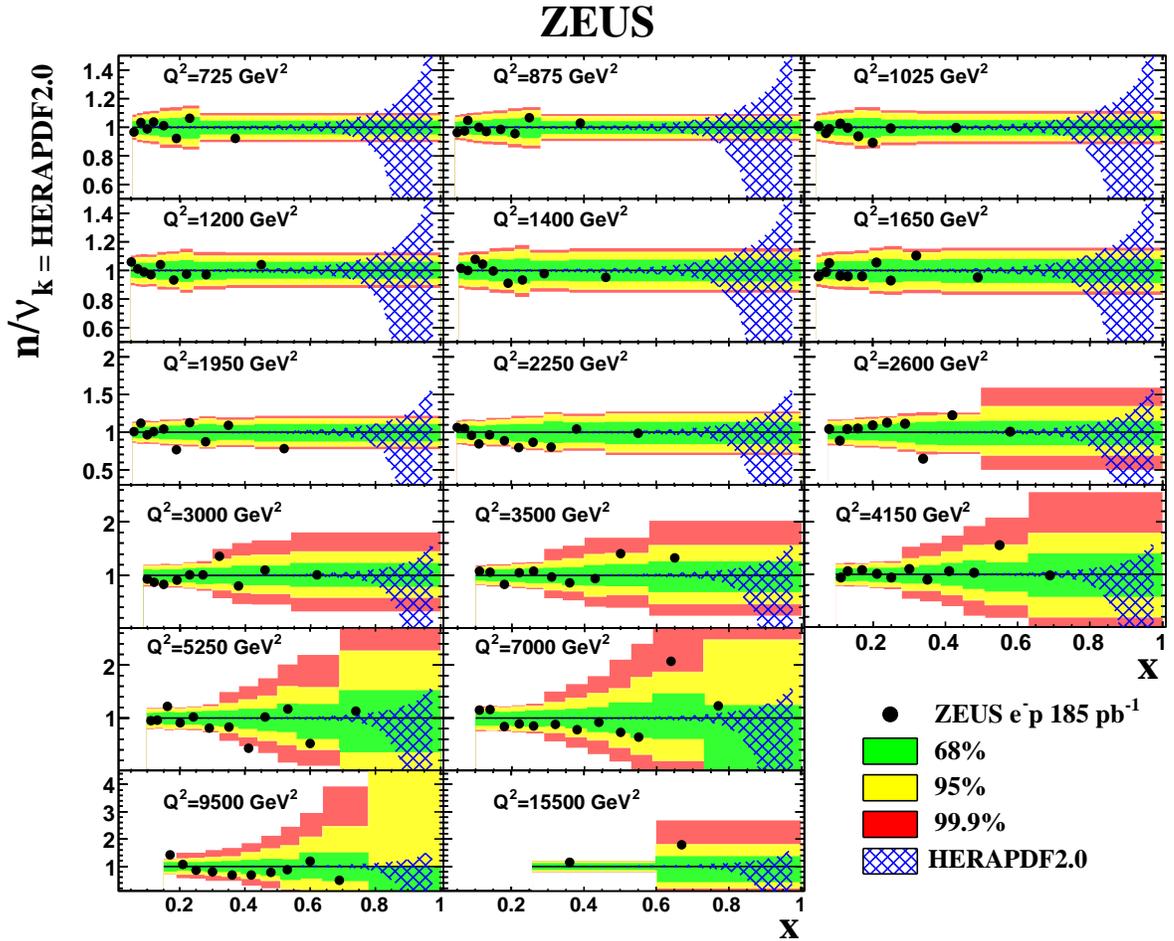}
\caption{Ratios of the number of observed events, $\boldsymbol n$, to the expectations from the HERAPDF2.0 set for $e^-p$ data. The points are plotted at the weighted average values of $x$. The green, yellow and red bands give the smallest intervals~\protect\cite{Aggarwal:2011aa} containing  68, 95, 99~$\%$ probability, respectively.  The ranges get wider as the number of expected events decreases. The cross-hatched band shows the uncertainty in the prediction associated with HERAPDF2.0.}
\label{fig:rec_events_bands_eMp}
\end{figure}

\begin{figure}
\centering
\includegraphics[scale=.8]{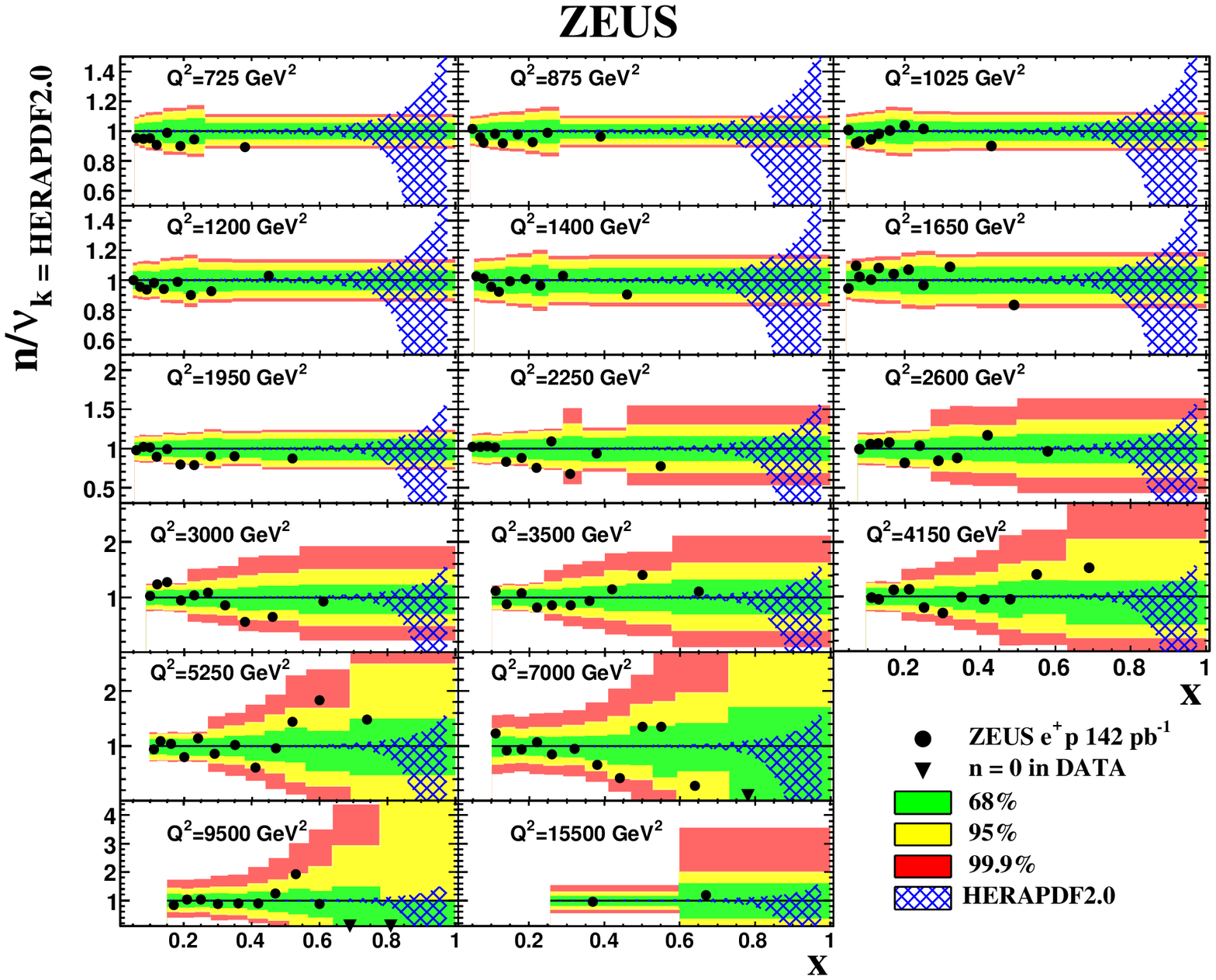}
\caption{Ratios of the number of observed events, $\boldsymbol n$, to the expectations from the HERAPDF2.0  set for $e^+p$ data.  The points are plotted at the weighted average values of $x$.The green, yellow and red bands give the smallest intervals~\protect\cite{Aggarwal:2011aa} containing  68, 95, 99~$\%$ probability, respectively.  The ranges get wider as the number of expected events decreases. The cross-hatched band shows the uncertainty in the prediction associated with HERAPDF2.0.}
\label{fig:rec_events_bands_ePp}
\end{figure}

Figures~\ref{fig:rec_events_bands_eMp} and~\ref{fig:rec_events_bands_ePp} present a comparison of the number of events observed in data and the predictions from HERAPDF2.0 PDF for  the $e^+p$ and $e^-p$ data, respectively.  The figures show that the data generally lie in the  68\% range of expected results.  However, for the $e^+p$ data, the HERAPDF2.0-based predictions are on average $3.1$\% higher than the data while for the $e^-p$ data, the HERAPDF2.0-based predictions are on average $0.9$\% higher than the data. Figures similar to Figs.~\ref{fig:rec_events_bands_eMp} and~\ref{fig:rec_events_bands_ePp} are available online for the PDF sets ABMP16, AMB11, NNPDF3.1, NNPDF2.3, CT14 and MMHT2014.

The overall probabilities for the observed data sets given the expectations from the different PDF sets were calculated using Eq.~(\ref{eq:probability}).  The natural logarithms of these probabilities are $-538.3$ and $-581.5$ for the $e^+p$ and $e^-p$ data sets, respectively, for HERAPDF2.0.  The corresponding $p$-values taking these probabilities as test statistics are $0.3$ and $0.03$. These probabilities account only for the statistical uncertainties. A summary of the results for different PDF sets is given in Table~\ref{tab:probs}.  The table gives the ratio of the probabilities found for different PDF sets (P1) to those found using HERAPDF2.0 (P2), see Eq.~(\ref{Eq:chsq}).  In addition, the effective $\Delta \chi^2$ as defined in Eq.~(\ref{Eq:chsq}) is provided.   


The predictions from all PDF sets yield $p$-values in an acceptable range for the $e^+p$ data, indicating good overall agreement with the observed data. There are nevertheless significant differences in the probabilities. The highest probability for the $e^+p$ data is found with the CT14 PDF set, with a Bayes Factor of $5.9 \cdot 10^5$ relative to HERAPDF2.0,  corresponding to $\Delta \chi^2=-27$.    For the $e^-p$ data set, large differences are seen in the $p$-values, with only the HERAPDF2.0 and the ABM11 and ABMP16 sets yielding sizeable $p$-values.  The Bayes Factors vary by more than $6$ orders of magnitude, resulting in $\Delta \chi^2$ values up to $31$.  
The bin-by-bin comparison of the nominal predictions for the different PDF sets to the the observations are given in the Appendix A.

The probability evaluation was carried out in two $x$ ranges: the `higher $x$' range is defined as the three highest-$x$ bins in each $Q^2$ range.  The remaining $x$ range is labeled `lower\,$x$'.  A substantial part of the data in the higher-$x$ range has not been used in PDF extractions, while the data in the lower-$x$ range have been included (albeit using different reconstruction techniques and coarser binning).  The results for the different $x$ ranges for both $e^+p$ and $e^-p$ data are given in Table~\ref{tab:Prob1}.  There are significant differences in both $x$ ranges.  The predictions from the ABM11 and ABMP16 sets yield lower probabilities than HERAPDF2.0 in the lower-$x$ range of the $e^-p$ data, but higher probabilities in the higher-$x$ range, while most of the difference in $\Delta \chi^2$ between CT14 and HERAPDF2.0 occurs in the higher-$x$ range. For the $e^+p$ data, similar values of $\Delta \chi^2$ are observed in both $x$ ranges. These results indicate that the high-$x$ data indeed have some discriminating power.

An effective statistical power of the higher-$x$ data was estimated as follows.  The normalization of the HERAPDF2.0 prediction was allowed to vary and an optimum was found using only the higher-$x$ data.  The change in the normalization resulting in a decrease of the logarithm of the likelihood of 0.5 was then determined. This normalization change was found to be small: $1.4$~\% for  the $e^+p$ data and $1.2$~\% for  the $e^-p$ data, indicating that these data should bring important new information.



All PDF sets provide not only nominal values for the parton distribution functions but also uncertainties that translate into uncertainties on the predicted cross sections.  The uncertainties for HERAPDF2.0  are provided as so-called variants; for all of these, predictions for the observed number of events were calculated.  The resulting spread in the predictions is typically much smaller than the difference seen between predictions from different PDF sets as discussed above. The probabilities to observe the data for the different variants of HERAPDF2.0 were evaluated, with Bayes Factors relative to the nominal predictions ranging from $0.4\rightarrow 3$, the largest of this range corresponding to  $\Delta \chi^2\approx 2$.  Given that the HERAPDF extraction uses a $\Delta \chi^2=1$ convention for evaluating uncertainties, a number of these variants would likely be excluded if the ZEUS high-$x$ data were included in the PDF extraction.

\section{Systematic uncertainties}
\label{sec:systematics}

In the discussion above, the nominal predictions from different PDF sets were used to predict event numbers, and a Poisson probability was calculated based on the observed number of events.  This procedure accounts only for statistical uncertainties.  In this section, the impact of systematic uncertainties on the probabilities is discussed.  Two classes of systematic uncertainties can be distinguished $-$ those that affect the predictions at the physics simulation level (the $\boldsymbol \mu$ and $\boldsymbol R$ values) and those that affect the matrix $\boldsymbol T$ accounting for detector and analysis effects.  These are discussed separately for HERAPDF2.0. 

For many of the PDF sets discussed in this paper, the combined H1 and ZEUS data~\cite{herapdf2.0} were included in their fitting procedure.  A large fraction of the ZEUS neutral current data~\cite{zeushiQ2paper,ZEUS2NCe} entering the combination was taken during the same running period as the high-$x$ data set considered here, and some of the systematic uncertainties are either identical to those discussed below or  correlated to them.  These uncertainties were already accounted for in the PDF extraction; i.e., the systematic uncertainties described below are not independent of the systematic uncertainties assigned to the PDF sets.  A new PDF extraction would be necessary to account properly for the correlated systematic uncertainties.  This point is discussed further below. 

\subsection{Uncertainties on  ${\boldsymbol \mu}$ and ${\boldsymbol R}$ }
Possible systematic effects on the predictions are due to imperfect values of ${\boldsymbol R}$ and a small net polarization of the electron and positron beams.  The latter effect was found to be negligible~\cite{highxpaper:1}.  To test whether the diagonal matrix for ${\boldsymbol R}$ determined using the CTEQ5D PDF set was sufficiently accurate, the ratio of the generator-level predictions to the Born-level predictions was compared for different PDF sets for all bins used in the analysis.  The ratios were found to be within the MC statistical uncertainties and typically well within $1$\%.  No systematic trends were observed.  Variations in the Born-level cross sections are therefore not expected to produce significant uncertainties in the determination of  ${\boldsymbol R}$.  Limitations could still be present due to missing effects in the MC simulations used in this analysis.  Should an improved simulation of radiative effects become available, the $\boldsymbol R$ matrix should be reevaluated and implemented.

The primary source of systematic uncertainty affecting ${\boldsymbol \mu_k}$ is the luminosity uncertainty, which acts as a scale factor on the elements of $\boldsymbol \mu_k$.  The effect of the luminosity uncertainty was evaluated by increasing and decreasing ${\boldsymbol \mu_k}$  by $\pm 1.8$\%, the  luminosity uncertainty of the data set~\cite{highxpaper:1}, separately for the $e^+p$ and $e^-p$ predictions, and recalculating the probabilities. The results are shown in Table~\ref{tab:lumi}. 

For $e^-p$ data, an upward shift of the luminosity improved the agreement between predictions from the CT14, MMHT2014, NNPDF2.3 and NNPDF3.1 sets and the data, while worsening the agreement for HERAPDF2.0 and the ABM11 and ABMP16 sets. The negative shift  gave lower (for the CT14, MMHT2014 and NNPDF2.3 and NNPDF3.1 much lower)  probabilities. For  HERAPDF2.0,  the probability was not significantly changed.

For the $e^+p$ data, the positive normalization shift made small changes for the CT14, MMHT2014, NNPDF2.3 and NNPDF3.1 sets and resulted in significantly worse probabilities for the HERAPDF2.0 and the ABM11 and ABMP16 sets. For the negative shift, the probability was significantly improved for HERAPDF2.0  and the ABM11 and ABMP16 sets while it generally became worse for the other PDF sets. Given that the normalization uncertainty is related to the data, it cannot be applied separately for the different PDFs sets.  This implies that  the significant difference in the predictions of the PDFs is not resolved by a change in the normalizations and no clear preference is seen for a given PDF set based strictly on the normalization of the data. 

\subsection{Uncertainties on ${\boldsymbol T}$}

One source of uncertainty on ${\boldsymbol T}$ is the loss of information due to the finite bin sizes in the kinematic variables.  A related source of uncertainty derives from using the same matrix for all PDF sets. To test the first effect, the bin sizes at the generator and Born level were systematically decreased until no further significant changes were observed in the predictions.  Both effects were then evaluated by replacing the matrix procedure with an event-by-event reweighting procedure.  The values of $\boldsymbol \nu_k$ were evaluated in this method as:
\begin{equation}
  \nu_{j,k} = \sum_m \frac{ \frac {d^2 \sigma(x,Q^{2}|\rm {PDF}_{\it{k}})}{dx dQ^2}  }{\frac {d^2 \sigma(x,Q^{2}|\rm {CTEQ5D})} {dx dQ^2}} \omega_{m}^{\rm{MC}}\omega_{m}^{\rm {sim}} I(m\in j) \; .
\label{eqn:nu_j_pdf}		
\end{equation}
 The sum runs over all generated MC events.  The indicator function is used to select events reconstructed in bin $j$; the ratio of differential cross sections is evaluated using the Born-level differential cross sections calculated from the kinematic quantities  at the generator level.  The numerator is the differential cross section for a desired PDF set $k$, while the denominator is the differential cross section from the CTEQ5D PDF set.  It was verified that the results are effectively identical for the different PDF sets considered in this paper.  The differences in $\boldsymbol \nu_k$ were typically at the level of $< 0.1$\% with a maximum difference of 1\% seen in the highest $x$ and $Q^2$ bins.


The limited size of the MC sample resulted in an uncertainty on ${\boldsymbol T}$. The statistical uncertainties on individual matrix elements are typically $\ll 1$\%, ranging to $1$\% in the highest $Q^2$ and $x$ bins.  
Propagating these uncertainties to the entries in  ${\boldsymbol \nu_k}$  results in negligible statistical uncertainties in the predicted numbers of events.


The variation in ${\boldsymbol T}$  from the dominant sources of uncertainty given in the ZEUS analysis~\cite{highxpaper:1} was further investigated. For each systematic effect considered,  ${\boldsymbol T}$ was reevaluated and used to produce a new set of predictions for the number of reconstructed events in the cross-section bins.  
The probabilities can change by up to a factor $10$, corresponding to effective $\chi^2$ changes of almost 5 units.  Although these changes are considerably smaller than those resulting from a change in the normalization, they should clearly be included in any new PDF extraction. 
\section{Prescription for PDF extractions including the high-$\boldsymbol x$ data}
Combining the ZEUS high-$x$ data with other HERA data in the extraction of PDFs will require some care. The details will depend on the procedure, the emphasis of the study, and whether the high-$x$ data are used in addition to the combined HERA data~\cite{herapdf2.0} or as a substitute for a part of the ZEUS data in a PDF extraction using individual ZEUS and H1 data sets. The latter $\it ansatz$  could be used to answer the question  whether and how the extended $x$ range and the finer binning of the high-$x$ data~\cite{highxpaper:1} impact the high-$x$ part of the PDFs. For the HERA II data with $\sqrt{s}=318$~GeV, the full set of H1 $e^+$ and $e^-$ neutral current data~\cite{H1hiQ2paper} could be used together with the standard ZEUS neutral current data~\cite{zeushiQ2paper,ZEUS2NCe} for $Q^2 < 650 $~GeV$^2$ and the ZEUS high-$x$ data~\cite{highxpaper:1} for $Q^2 \geq 650$~GeV$^2$. The standard and high-$x$ ZEUS data have a completely correlated normalization uncertainty while most of the remaining systematic uncertainties can be considered uncorrelated, since the reconstruction methods were significantly different. If wanted, the individual HERA II data sets could be augmented with the combined HERA I data~\cite{HERAIcombi} and HERA II combined data sets for lower  $\sqrt s$. However, these data are not expected to have a large impact on the high-$x$ part of the PDFs.

The use of the high-$x$ data in addition to the combined HERA data~\cite{herapdf2.0} may however be preferred, since this permits a coherent treatment of systematic uncertainties over as much of the phase space as possible. Care has to be taken to avoid double counting  by ensuring that bins individually added do not overlap with the bins from the remaining combined data.  Table C.1 provides one possible selection. Combined $e^+$ and $e^-$ data~\cite{herapdf2.0} for $x=0.4$ and $x=0.6$ were removed for $Q^2$ values at which ZEUS data~\cite{zeushiQ2paper,ZEUS2NCe} contributed. The finer binned ZEUS high-$x$ data including the integrated bins were added as also listed in Table~C.1. The individual $e^+$ and $e^-$ H1 data points~\cite{H1hiQ2paper} for the $x$ and $Q^2$ values of the removed combined data points should also be added back individually.

For the procedure using combined and individual data, the normalizations of the individual data sets have to be adjusted. During the combination of H1 and ZEUS data sets, the normalizations of all individual data sets were shifted~\cite{herapdf2.0}. The normalization shifts of the ZEUS $e^+$ and $e^-$ data sets as applied in the combination have to be applied to the individual ZEUS NC high-$x$ $e^+$ and $e^-$ data sets. This implies a $\it reduction$ in the luminosity by the factors given in Table~C.2 for these data sets. For the individual NC $e^+$ and $e^-$ H1 data points~\cite{H1hiQ2paper} to be reintroduced in the analysis, the cross sections have to be increased by the factors given in Table~C.2. The normalization uncertainties quoted for the individual data sets should be treated as correlated systematic uncertainties on the individual ZEUS and H1 data points. The remaining systematic uncertainties should be treated as uncorrelated due to the different reconstruction techniques.

Different selections from the choices as listed in Table~C.1 are conceivable. Parameterizations with sufficient flexibility in the high-$x$ region could benefit from a more extended use of the high-$x$ data. In any case, the procedures using Poisson statistics as outlined in this paper should be employed to make the optimal use of the high-$x$ data.
\section{Conclusions}
The ZEUS high-$x$ data are unique and, to-date, have not been used in the extraction of proton parton distribution functions. They should be included in future PDF extractions using the matrix approach described in this paper.  This data set will give access to a previously unused kinematic region, and will help constrain the large uncertainties on the partonic structure at the highest values of $x$. 

The comparison of predictions from modern PDF sets in the kinematic range covered by the ZEUS high-$x$ data show large differences that are well beyond the uncertainties associated with the predictions, indicating that the uncertainties on the PDFs are underestimated.  At the highest values of $x$, they could be significantly larger than currently thought, making the usage of high-$x$ data even more important. The quantitative effect of the ZEUS high-$x$ data on reducing the uncertainties is difficult to ascertain without carrying out the PDF extraction procedure in full detail.  A proposal for including these data in future PDF extractions has been outlined.
\clearpage

\section*{Acknowledgements}
\label{sec-ack}

\Zacknowledge

\clearpage
{
\ifzeusbst
  \ifzmcite
     \bibliographystyle{./BiBTeX/bst/l4z_default3}
  \else
     \bibliographystyle{./BiBTeX/bst/l4z_default3_nomcite}
  \fi
\fi
\ifzdrftbst
  \ifzmcite
    \bibliographystyle{./BiBTeX/bst/l4z_draft3}
  \else
    \bibliographystyle{./BiBTeX/bst/l4z_draft3_nomcite}
  \fi
\fi
\ifzbstepj
  \ifzmcite
    \bibliographystyle{./BiBTeX/bst/l4z_epj3}
  \else
    \bibliographystyle{./BiBTeX/bst/l4z_epj3_nomcite}
  \fi
\fi
\ifzbstjhep
  \ifzmcite
    \bibliographystyle{./BiBTeX/bst/l4z_jhep3}
  \else
    \bibliographystyle{./BiBTeX/bst/l4z_jhep3_nomcite}
  \fi
\fi
\ifzbstnp
  \ifzmcite
    \bibliographystyle{./BiBTeX/bst/l4z_np3}
  \else
    \bibliographystyle{./BiBTeX/bst/l4z_np3_nomcite}
  \fi
\fi
\ifzbstpl
  \ifzmcite
    \bibliographystyle{./BiBTeX/bst/l4z_pl3}
  \else
    \bibliographystyle{./BiBTeX/bst/l4z_pl3_nomcite}
  \fi
\fi
{\raggedright
\bibliography{./syn.bib,%
              ./myref.bib,%
              ./BiBTeX/bib/l4z_zeus.bib,%
              ./BiBTeX/bib/l4z_h1.bib,%
              ./BiBTeX/bib/l4z_articles.bib,%
              ./BiBTeX/bib/l4z_books.bib,%
              ./BiBTeX/bib/l4z_conferences.bib,%
              ./BiBTeX/bib/l4z_misc.bib,%
              ./BiBTeX/bib/l4z_preprints.bib}}
}
\vfill\eject

\begin{table}
\centering
\begin{tabular}{|c|ccc|ccc|}
\hline
 &  \multicolumn{3}{c|}{$e^-p$} & \multicolumn{3}{|c|}{$e^+p$}  \\
\hline
 PDF &   $p$-value&P1/P2 & $\Delta \chi^2 $& $p$-value&P1/P2 &  $\Delta \chi^2 $\\
\hline
  HERAPDF2.0  & $ 2.8\times10^{-2} $ & $ 1.0  $ & $ 0.0 $ & $ 0.35 $  & $ 1.0  $ & $ 0.0 $\\ 
  CT14  & $ 3.2 \times 10^{-3} $ & $ 7.6 \times 10^{-3} $ & $ 9.8 $ & $ 0.82 $  & $ 5.9 \times 10^{+5} $ & $ -27 $\\ 
  MMHT2014  & $ 2.3 \times 10^{-3} $ & $ 2.1 \times 10^{-3} $ & $ 12 $ & $ 0.82 $  & $ 4.7 \times 10^{+5} $ & $ -26 $\\ 
  NNPDF3.1  & $ 3.9 \times 10^{-4} $ & $ 3.2 \times 10^{-6} $ & $ 25 $ & $ 0.73 $  & $ 9.0 \times 10^{+4} $ & $ -23 $\\ 
  NNPDF2.3  & $ 1.3 \times 10^{-4} $ & $ 2.3 \times 10^{-7} $ & $ 31 $ & $ 0.70 $  & $ 4.2 \times 10^{+4} $ & $ -21 $\\ 
  ABMP16  & $ 2.6 \times 10^{-2} $ & $ 9.0 \times 10^{-1} $ & $ 0.21 $ & $ 0.64 $  & $ 6.1 \times 10^{+2} $ & $ -13 $\\ 
  ABM11  & $ 3.3 \times 10^{-2} $ & $ 7.2 \times 10^{-1} $ & $ 0.67 $ & $ 0.45 $  & $ 2.8  $ & $ -2.1 $\\

\hline
\end{tabular}
\caption{The  $p$-values, Bayes Factors (P1/P2) and $\Delta\chi^2$ from comparisons of predictions using different NNLO PDF sets  to the observed numbers of events.  The Bayes Factor is calculated relative to HERAPDF2.0, as is $\Delta \chi^2$.   The results are shown separately for the $e^-p$ and $e^+p$ data sets.}
\label{tab:probs}
\end{table}

\begin{table}
\centering

\footnotesize
\begin{tabular}{|c|cc|cc||cc|cc|}
\hline
             & \multicolumn{4}{c||}{$e^-p$} & \multicolumn{4}{c|}{$e^+p$}  \\
\hline
 &\multicolumn{2}{c|}{lower $x$} & \multicolumn{2}{c||}{higher $x$}& \multicolumn{2}{c|}{lower $x$} & \multicolumn{2}{c|}{higher $x$} \\

 \large{PDF} &  P1/P2 & $\Delta \chi^2 $&  P1/P2 & $\Delta \chi^2 $&  P1/P2 & $\Delta \chi^2 $ & P1/P2 & $\Delta \chi^2 $\\
\hline
\hline
  HERAPDF2.0  & $ 1.0 $ & $ 0.0 $ & $ 1.0  $ & $ 0.0 $ & $  1.0  $ & $ 0.0 $&$ 1.0  $&$ 0.0  $\\ 
   CT14  & $ 6.0\times10^{-1} $ & $ 1.0 $ & $ 1.3\times10^{-2}  $ & $ 8.7 $ & $  1.6\times10^{+3}  $ & $ -15 $&$ 3.6\times10^{+2}  $&$ -12 $\\ 
  MMHT2014  & $ 7.1\times10^{-2} $ & $ 5.3 $ & $ 2.9\times10^{-2}  $ & $ 7.1 $ & $  1.3\times10^{+3}  $ & $ -14 $&$ 3.7\times10^{+2}  $&$ -12 $\\
  NNPDF3.1  & $ 9.1\times10^{-5} $ & $ 19 $ & $ 3.5\times10^{-2}  $ & $ 6.7 $ & $  2.5\times10^{+2}  $ & $ -11 $&$ 3.6\times10^{+2}  $&$ -12  $\\ 
  NNPDF2.3  & $ 8.0\times10^{-6} $ & $ 23 $ & $ 2.9\times10^{-2}  $ & $ 7.1 $ & $  1.2\times10^{+2}  $ & $ -9.5 $&$ 3.7\times10^{+2}  $&$ -12  $\\  
  ABMP16  & $ 2.3\times10^{-1} $ & $ 3.0 $ & $ 4.0  $ & $ -2.7 $ & $  4.8\times10^{+1}  $ & $ -7.8 $&$ 1.3\times10^{+1}  $&$ -5.1  $\\ 
  ABM11  & $ 2.3\times10^{-1} $ & $ 3.0 $ & $ 3.2  $ & $ -2.3 $ & $  4.2  $ & $ -2.9 $&$ 6.7\times10^{-1}  $&$ 0.8  $\\

\hline
\end{tabular}
\caption{ The Bayes Factor (P1/P2) and $\Delta\chi^2$ (calculated relative to HERAPDF2.0) from comparisons of predictions using different NNLO PDF  sets to the observed numbers of events are shown for two different $x$ ranges.
The `higher-$x$' region is defined as the highest three $x$ bins in each $Q^2$ range.  The remaining $x$ range is labeled `lower-$x$'.}
\label{tab:Prob1}
\end{table}


\clearpage
%
\appendix

\refstepcounter{section}
\section*{Appendix A}
\label{appendix}
\setcounter{table}{0}
\renewcommand\thetable{\thesection.\arabic{table}}

\begin{landscape}
\begin{center}
\footnotesize


\tablefirsthead{
\multicolumn{11}{l}{
{\normalsize}}\\
\hline
{$Q^2$} & 
\multicolumn{1}{c|}{$x$} & 
Data& 
\multicolumn{2}{c|}{HERAPDF2.0}  &
\multicolumn{2}{c|}{CT14}  &
\multicolumn{2}{c|}{MMHT2014}  &
\multicolumn{2}{c|}{NNPDF3.1}  &
\multicolumn{2}{c|}{NNPDF2.3}  &
\multicolumn{2}{c|}{ABMP16} &
\multicolumn{2}{c|}{AMB11}   \\
\hline
{(GeV$^2$)} & 
$$ &
$\boldsymbol{n}$ &
 $\boldsymbol{\nu}$ &
$P(\boldsymbol{n}|\boldsymbol{\nu})$ &
$\boldsymbol{\nu}$ &
$P(\boldsymbol{n}|\boldsymbol{\nu})$ &
$\boldsymbol{\nu}$ &
$P(\boldsymbol{n}|\boldsymbol{\nu})$ &
$\boldsymbol{\nu}$ &
$P(\boldsymbol{n}|\boldsymbol{\nu})$   & 
$\boldsymbol{\nu}$ &
$P(\boldsymbol{n}|\boldsymbol{\nu})$ &
$\boldsymbol{\nu}$ &
$P(\boldsymbol{n}|\boldsymbol{\nu})$ &
$\boldsymbol{\nu}$ &
$P(\boldsymbol{n}|\boldsymbol{\nu})$\\
\hline \hline}
\tablelasttail{
\hline}

\renewcommand{\arraystretch}{1.2}

\tablehead{
\multicolumn{11}{l}{
{\normalsize {\bf Table \thetable\ } (continued):}}\\
\hline
{$Q^2$} & 
\multicolumn{1}{c|}{$x$} & 
Data& 
\multicolumn{2}{c|}{HERAPDF2.0}  &
\multicolumn{2}{c|}{CT14}  &
\multicolumn{2}{c|}{MMHT2014}  &
\multicolumn{2}{c|}{NNPDF3.1}  &
\multicolumn{2}{c|}{NNPDF2.3}  &
\multicolumn{2}{c|}{ABMP16} &
\multicolumn{2}{c|}{AMB11}   \\
\hline
{(GeV$^2$)} & 
$$ &
$\boldsymbol{n}$ &
 $\boldsymbol{\nu}$ &
$P(\boldsymbol{n}|\boldsymbol{\nu})$ &
$\boldsymbol{\nu}$ &
$P(\boldsymbol{n}|\boldsymbol{\nu})$ &
$\boldsymbol{\nu}$ &
$P(\boldsymbol{n}|\boldsymbol{\nu})$ &
$\boldsymbol{\nu}$ &
$P(\boldsymbol{n}|\boldsymbol{\nu})$   & 
$\boldsymbol{\nu}$ &
$P(\boldsymbol{n}|\boldsymbol{\nu})$ &
$\boldsymbol{\nu}$ &
$P(\boldsymbol{n}|\boldsymbol{\nu})$ &
$\boldsymbol{\nu}$ &
$P(\boldsymbol{n}|\boldsymbol{\nu})$\\
\hline \hline}
\tabletail{
\hline
}
\tabletail{
\hline
}
\tablelasttail{\hline}
\footnotesize
\topcaption{
 The comparison of event counts in data and in MC for the $e^-p$ sample, using different PDFs.
    The first two columns of the table contain the $Q^2$ and $x$ values
    for the center of the bin and the third column contains the number of events reconstructed in
    the bin in data $(\boldsymbol n)$. The further columns contain expectations $(\boldsymbol \nu)$ and the probability $P(\boldsymbol {n}|\boldsymbol{\nu})$ for the PDFs discussed in the paper.
    } 
\begin{supertabular}{|c|c|c|c|c|c|c|c|c|c|c|c|c|c|c|c|c|c|}
$ 725 $ & $ 0.06 $ & $ 743         $ & $ 768.1 $ & $ 0.010 $ & $ 751.6 $ & $ 0.014 $ & $ 747.6 $ & $ 0.014 $ & $ 736.2 $ & $ 0.014 $ & $ 732.5 $ & $ 0.014 $ & $ 750.0 $ & $ 0.014 $ & $ 759.3 $ & $ 0.012 $ \\ 
$ 725 $ & $ 0.08 $ & $ 580         $ & $ 560.9 $ & $ 0.012 $ & $ 547.2 $ & $ 0.006 $ & $ 545.1 $ & $ 0.006 $ & $ 537.1 $ & $ 0.003 $ & $ 535.0 $ & $ 0.003 $ & $ 549.7 $ & $ 0.007 $ & $ 555.6 $ & $ 0.010 $ \\ 
$ 725 $ & $ 0.10 $ & $ 441         $ & $ 445.0 $ & $ 0.019 $ & $ 433.7 $ & $ 0.018 $ & $ 432.3 $ & $ 0.017 $ & $ 426.6 $ & $ 0.015 $ & $ 425.3 $ & $ 0.014 $ & $ 438.0 $ & $ 0.019 $ & $ 442.0 $ & $ 0.019 $ \\ 
$ 725 $ & $ 0.12 $ & $ 416         $ & $ 401.2 $ & $ 0.015 $ & $ 391.0 $ & $ 0.009 $ & $ 390.0 $ & $ 0.008 $ & $ 385.9 $ & $ 0.006 $ & $ 384.7 $ & $ 0.006 $ & $ 396.1 $ & $ 0.012 $ & $ 399.4 $ & $ 0.014 $ \\ 
$ 725 $ & $ 0.16 $ & $ 283         $ & $ 279.9 $ & $ 0.023 $ & $ 273.0 $ & $ 0.020 $ & $ 272.3 $ & $ 0.019 $ & $ 270.3 $ & $ 0.018 $ & $ 269.3 $ & $ 0.017 $ & $ 277.1 $ & $ 0.022 $ & $ 279.4 $ & $ 0.023 $ \\ 
$ 725 $ & $ 0.19 $ & $ 248         $ & $ 268.4 $ & $ 0.011 $ & $ 261.4 $ & $ 0.018 $ & $ 260.9 $ & $ 0.018 $ & $ 259.4 $ & $ 0.020 $ & $ 258.4 $ & $ 0.021 $ & $ 265.7 $ & $ 0.014 $ & $ 267.9 $ & $ 0.012 $ \\ 
$ 725 $ & $ 0.23 $ & $ 227         $ & $ 213.3 $ & $ 0.017 $ & $ 206.7 $ & $ 0.010 $ & $ 206.5 $ & $ 0.010 $ & $ 205.5 $ & $ 0.009 $ & $ 204.8 $ & $ 0.008 $ & $ 210.7 $ & $ 0.014 $ & $ 212.0 $ & $ 0.016 $ \\ 
$ 725 $ & $ 0.63 $ & $ 504         $ & $ 545.5 $ & $ 0.004 $ & $ 516.1 $ & $ 0.015 $ & $ 517.8 $ & $ 0.015 $ & $ 517.2 $ & $ 0.015 $ & $ 516.5 $ & $ 0.015 $ & $ 535.0 $ & $ 0.007 $ & $ 542.5 $ & $ 0.004 $ \\
\hline
\end{supertabular}
\end{center}

\end{landscape}

\clearpage
\begin{landscape}
\begin{center}
\footnotesize


\tablefirsthead{
\multicolumn{11}{l}{
{\small}}\\
\hline
{$Q^2$} & 
\multicolumn{1}{c|}{$x$} & 
Data& 
\multicolumn{2}{c|}{HERAPDF2.0}  &
\multicolumn{2}{c|}{CT14}  &
\multicolumn{2}{c|}{MMHT2014}  &
\multicolumn{2}{c|}{NNPDF3.1}  &
\multicolumn{2}{c|}{NNPDF2.3}  &
\multicolumn{2}{c|}{ABMP16} &
\multicolumn{2}{c|}{AMB11}   \\
\hline
{(GeV$^2$)} & 
$$ &
$\boldsymbol{n}$ &
 $\boldsymbol{\nu}$ &
$P(\boldsymbol{n}|\boldsymbol{\nu})$ &
$\boldsymbol{\nu}$ &
$P(\boldsymbol{n}|\boldsymbol{\nu})$ &
$\boldsymbol{\nu}$ &
$P(\boldsymbol{n}|\boldsymbol{\nu})$ &
$\boldsymbol{\nu}$ &
$P(\boldsymbol{n}|\boldsymbol{\nu})$   & 
$\boldsymbol{\nu}$ &
$P(\boldsymbol{n}|\boldsymbol{\nu})$ &
$\boldsymbol{\nu}$ &
$P(\boldsymbol{n}|\boldsymbol{\nu})$ &
$\boldsymbol{\nu}$ &
$P(\boldsymbol{n}|\boldsymbol{\nu})$\\
\hline \hline}
\tablelasttail{
\hline}

\tablehead{
\multicolumn{11}{l}{
{\small {\bf Table \thetable\ } (continued):}}\\
\hline
{$Q^2$} & 
\multicolumn{1}{c|}{$x$} & 
Data& 
\multicolumn{2}{c|}{HERAPDF2.0}  &
\multicolumn{2}{c|}{CT14}  &
\multicolumn{2}{c|}{MMHT2014}  &
\multicolumn{2}{c|}{NNPDF3.1}  &
\multicolumn{2}{c|}{NNPDF2.3}  &
\multicolumn{2}{c|}{ABMP16} &
\multicolumn{2}{c|}{AMB11}   \\
\hline
{(GeV$^2$)} & 
$$ &
$\boldsymbol{n}$ &
 $\boldsymbol{\nu}$ &
$P(\boldsymbol{n}|\boldsymbol{\nu})$ &
$\boldsymbol{\nu}$ &
$P(\boldsymbol{n}|\boldsymbol{\nu})$ &
$\boldsymbol{\nu}$ &
$P(\boldsymbol{n}|\boldsymbol{\nu})$ &
$\boldsymbol{\nu}$ &
$P(\boldsymbol{n}|\boldsymbol{\nu})$   & 
$\boldsymbol{\nu}$ &
$P(\boldsymbol{n}|\boldsymbol{\nu})$ &
$\boldsymbol{\nu}$ &
$P(\boldsymbol{n}|\boldsymbol{\nu})$ &
$\boldsymbol{\nu}$ &
$P(\boldsymbol{n}|\boldsymbol{\nu})$\\
\hline \hline}
\tabletail{
\hline
}
\tabletail{
\hline
}
\tablelasttail{\hline}
\footnotesize

\topcaption{
    The comparison of event counts in data and in MC for the $e^+p$ sample, using different PDFs.
    The first two columns of the table contain the $Q^2$ and $x$ values
    for the center of the bin and the third column contains the number of events reconstructed in
    the bin in data $(\boldsymbol n)$. The further columns contain expectations $(\boldsymbol \nu)$ and the probability  $P(\boldsymbol {n}|\boldsymbol{\nu})$  for the PDFs discussed in the paper.
    }

\begin{supertabular}{|c|c|c|c|c|c|c|c|c|c|c|c|c|c|c|c|c|c|}
$ 725 $ & $ 0.06 $ & $ 567         $ & $ 594.3 $ & $ 0.009 $ & $ 581.8 $ & $ 0.014 $ & $ 578.8 $ & $ 0.015 $ & $ 569.9 $ & $ 0.017 $ & $ 566.9 $ & $ 0.017 $ & $ 580.5 $ & $ 0.014 $ & $ 587.4 $ & $ 0.012 $ \\ 
$ 725 $ & $ 0.08 $ & $ 418         $ & $ 440.2 $ & $ 0.011 $ & $ 429.6 $ & $ 0.017 $ & $ 428.1 $ & $ 0.017 $ & $ 421.6 $ & $ 0.019 $ & $ 419.9 $ & $ 0.019 $ & $ 431.6 $ & $ 0.016 $ & $ 435.9 $ & $ 0.013 $ \\ 
$ 725 $ & $ 0.10 $ & $ 321         $ & $ 337.2 $ & $ 0.015 $ & $ 328.7 $ & $ 0.020 $ & $ 327.8 $ & $ 0.021 $ & $ 323.4 $ & $ 0.022 $ & $ 322.3 $ & $ 0.022 $ & $ 332.0 $ & $ 0.018 $ & $ 334.9 $ & $ 0.017 $ \\ 
$ 725 $ & $ 0.12 $ & $ 274         $ & $ 302.3 $ & $ 0.006 $ & $ 294.7 $ & $ 0.011 $ & $ 294.0 $ & $ 0.012 $ & $ 290.9 $ & $ 0.015 $ & $ 290.0 $ & $ 0.015 $ & $ 298.6 $ & $ 0.009 $ & $ 301.0 $ & $ 0.007 $ \\ 
$ 725 $ & $ 0.16 $ & $ 207         $ & $ 209.1 $ & $ 0.027 $ & $ 204.0 $ & $ 0.027 $ & $ 203.5 $ & $ 0.027 $ & $ 201.9 $ & $ 0.026 $ & $ 201.2 $ & $ 0.025 $ & $ 207.1 $ & $ 0.028 $ & $ 208.7 $ & $ 0.028 $ \\ 
$ 725 $ & $ 0.19 $ & $ 176         $ & $ 195.2 $ & $ 0.011 $ & $ 190.2 $ & $ 0.018 $ & $ 189.8 $ & $ 0.018 $ & $ 188.7 $ & $ 0.019 $ & $ 187.9 $ & $ 0.020 $ & $ 193.3 $ & $ 0.013 $ & $ 194.8 $ & $ 0.012 $ \\ 
$ 725 $ & $ 0.23 $ & $ 152         $ & $ 161.0 $ & $ 0.025 $ & $ 156.0 $ & $ 0.031 $ & $ 155.9 $ & $ 0.031 $ & $ 155.1 $ & $ 0.031 $ & $ 154.6 $ & $ 0.032 $ & $ 159.1 $ & $ 0.028 $ & $ 160.0 $ & $ 0.026 $ \\ 
$ 725 $ & $ 0.63 $ & $ 371         $ & $ 415.4 $ & $ 0.002 $ & $ 393.1 $ & $ 0.011 $ & $ 394.5 $ & $ 0.010 $ & $ 394.0 $ & $ 0.011 $ & $ 393.5 $ & $ 0.011 $ & $ 407.5 $ & $ 0.004 $ & $ 413.2 $ & $ 0.002 $ \\
\hline
\end{supertabular}
\end{center}
\end{landscape}

\clearpage

\refstepcounter{section}
\section*{Appendix B}
\label{appendix-b}
\setcounter{table}{0}
\renewcommand\thetable{\thesection.\arabic{table}}

\begin{table}[!htbp]
\centering 
\begin{tabular}{|c|cc|cc|cc|cc|}
\hline
              & \multicolumn{4}{c|}{$e^-p$} & \multicolumn{4}{c|}{$e^+p$}  \\
\hline
PDF& \multicolumn{2}{c|}{lower $x$} & \multicolumn{2}{c|}{higher $x$}& \multicolumn{2}{c|}{lower $x$} & \multicolumn{2}{c|}{higher $x$} \\\hline

& P1/P2 & $\Delta\chi^2$& P1/P2 & $\Delta\chi^2$& P1/P2 & $\Delta\chi^2$& P1/P2 & $\Delta\chi^2$\\ \hline
\multicolumn{9}{|c|}{$+1.8$~\%} \\\hline
  HERAPDF2.0  & $ 1.7\times 10^{-3} $ & $ 13  $ & $ 0.08  $ & $ 5.0  $ & $ 8.6\times 10^{-5}  $ & $ 19 $&$ 0.01  $&$ 9.8  $\\ 
  CT14  & $ 28 $ & $ -6.7  $ & $ 16  $ & $ -5.5  $ & $ 5.0\times 10^{-2}  $ & $ 6.0 $&$ 0.37  $&$ 2.0  $\\ 
  MMHT2014  & $ 1.1\times 10^{+2} $ & $ -9.5  $ & $ 13  $ & $ -5.1  $ & $ 0.11  $ & $ 4.4 $&$ 0.31  $&$ 2.3 $\\  
  NNPDF3.1  & $ 4.9\times 10^{+3} $ & $ -17  $ & $ 15  $ & $ -5.4  $ & $ 2.2  $ & $ -1.5 $&$ 0.36  $&$ 2.0  $\\
\hline
\end{tabular}
\caption{The results from comparisons of predictions using different PDF sets increased by $1.8$~\% (top) and decreased by $1.8$~\% (bottom) with respect to the observed numbers of events.  The  Bayes Factor (P1/P2) and $\Delta\chi^2$ (calculated relative to the respective values for the given PDFs as given in Table~\ref{tab:Prob1}) are  shown  for the two different $x$ ranges defined in the text for the $e^-p$ and $e^+p$ data sets.}
\label{tab:lumi}
\end{table}

\clearpage

\refstepcounter{section}
\section*{Appendix C}
\label{appendix-b}
\setcounter{table}{0}
\renewcommand\thetable{\thesection.\arabic{table}}

\begin{table}[!htbp]
\centering 
\begin{tabular}{|c|c|}

\hline
\multicolumn{2}{|c|}{HERA Combined points to be removed} \\
\hline
$x$ & $Q^2$ (GeV$^{\,2}$) \\
\hline
0.4 & 1200 1500 2000 3000 20000\\
\hline
0.6 & 3000 8000\\
\hline
\hline
\multicolumn{2}{|c|}{ZEUS high-$x$ data to be added} \\
\hline
$x$ & $Q^2$ (GeV$^{\,2}$)\\
\hline
Integrated &  725 875 1025 1200 1400 \\
 & 1650 1950 2250 2600 3000\\
 & 3500 4150 5250 7000 9500 \\
\hline
0.37 & 1950 \\
\hline
0.4 & 2250\\
\hline
0.35, 0.44 &2600\\
\hline
0.39, 0.48 & 3000\\
\hline
0.57 & 4150 \\
\hline
0.53, 0.62 &5250\\
\hline
0.56, 0.66 & 7000\\
\hline
0.54, 0.61, 0.71 & 9500 \\
\hline
0.8 & 15500\\
\hline

\end{tabular}

\caption{List of $e^\pm p$ combined-data points~\protect\cite{herapdf2.0} to be removed and the ZEUS high-$x$ data points~\protect\cite{highxpaper:1} to be added, see Section~6 for details.}
\label{tab:dataPoints1}
\end{table}

\begin{table}[!htbp]
\centering 
\begin{tabular}{|c|c|c|}

\hline
  & $e^-p$ &$e^+p$\\
\hline
ZEUS Luminosity reduction factor& 0.987 & 0.975 \\
\hline
H1 cross section enhancement factor & 1.024 & 1.013 \\
\hline

\end{tabular}

\caption{Luminosity reduction factors for ZEUS high-x data and cross section enhancement factors for H1 data to adjust to normalization shifts applied during data combination.}
\label{tab:dataPoints2}
\end{table}
%
%
\end{document}